\documentclass[seceq]{ptptex}

\usepackage{graphicx,amsmath}
\usepackage{dcolumn}
\usepackage{bm}

\newcommand{\delr}{\partial_r}
\newcommand{\delt}{\partial_t}
\newcommand{\tb}{t_{\rm B}}
\newcommand{\da}{D_{\rm A}}

\newcommand{\tilC}{\widetilde C}
\newcommand{\tilD}{\widetilde D}
\newcommand{\tilE}{\widetilde E}

\newcommand{\tilG}{\widetilde G}
\newcommand{\tilH}{\widetilde H}
\newcommand{\tilI}{\widetilde I}

\preprintnumber[5cm]{
YITP-10-80\\
}

\title{
A Note on the Inverse Problem with LTB Universes}

\author{
Chul-Moon \textsc{Yoo}%
}
\inst{
Yukawa Institute for Theoretical Physics, \\
Kyoto University,
Kyoto 606-8502, Japan

\recdate{
July 15, 2010}

}
\abst{
The inverse problem with Lema\^itre-Tolman-Bondi (LTB) universe models 
is discussed. 
The LTB solution for the Einstein equations describes 
the spherically symmetric dust-filled spacetime. 
The LTB solution has two physical functional degrees of freedom
of the radial coordinate. 
The inverse problem is 
constructing an LTB model requiring that the 
LTB model be consistent with 
selected important observational data.  
In this paper, we assume that the observer is at the center and 
consider the distance-redshift relation $\da$ and 
the redshift-space mass density $\mu$ 
as the selected important observational data. 
We give $\da$ and $\mu$ as functions of the redshift $z$. 
Then,
we explicitly show that, 
for general functional forms of $\da(z)$ and $\mu(z)$, 
the regular solution does not necessarily exist 
in the whole redshift domain. 
We also show that the condition for the existence of the regular 
solution 
 is satisfied 
by the distance-redshift relation and 
the redshift-space mass density in $\Lambda$CDM models. 
Deriving regular differential equations for the inverse problem 
with the distance-redshift relation and the redshift-space mass density 
in $\Lambda$CDM models, we numerically solve them 
for the case $(\Omega_{\rm M0},\Omega_{\Lambda0})=(0.3,0.7)$. 
A set of analytic fitting functions for the resultant LTB universe model is given. 
How to solve the inverse problem with the simultaneous big-bang and 
a given function $\da(z)$ for the distance-redshift relation 
is provided in the Appendix. 
}

\begin{document}
\maketitle

\section{Introduction}
\label{sec:intro}

Anti-Copernican universe models are 
widely discussed in recent years as 
alternatives to the standard homogeneous cosmology 
with dark energy components. 
One of the simplest ways to construct an anti-Copernican model 
is to solve the inverse problem, 
which is the
reconstruction of the universe model from 
observational data. 
Although the isotropy of the universe around us has been confirmed
with high accuracy by the observation of the cosmic microwave 
background (CMB), this does not automatically imply
homogeneity of the universe.
Thus, in solving the inverse problem, we may assume
that the universe is spherically symmetric around us.
In addition, we usually assume that the universe is
dominated by cold dark matter, that is, by dust. 
The spherically symmetric dust-filled spacetime is described by the 
Lema\^itre-Tolman-Bondi (LTB) solution\cite{Lemaitre:1933gd,Tolman:1934za,Bondi:1947av}. 
The LTB solution
has three arbitrary functions of the radial coordinate
approximately corresponding to the density profile, 
the spatial curvature and the big-bang time perturbation,
with one of them being a gauge degree of freedom representing 
the choice of the radial coordinate.
These arbitrary functions may be determined by requiring that
the resulting LTB universe be consistent with selected important 
observational data (e.g., the distance-redshift relation). 
However, we should note that it is not apparent at all if these three 
functions have sufficient degrees of freedom to fit all of the important
observational data.

In the inverse problem, two approaches 
have been mainly considered. 
One is proposed by Mustapha, Hellaby and Ellis\cite{Mustapha:1998jb} 
in 1998. 
In this paper, we refer to this work and the approach used there 
as MHE and the MHE approach, respectively. 
In MHE, the angular diameter distance $\da$ and 
the redshift-space mass density $\mu$ are given as 
functions of the redshift $z$
to fix two physical 
functional degrees of freedom in LTB universe models. 

Recently, many authors have succeeded in 
solving the inverse problem by the MHE approach 
\cite{Lu:2007gr,McClure:2007hy,Celerier:2009sv,Kolb:2009hn,Dunsby:2010ts}, 
and constructed the LTB universe model 
whose distance-redshift relation and the redshift-space mass density 
agree with those in the concordance $\Lambda$CDM model. 
The other approach is proposed by Iguchi, Nakamura and Nakao 
in 2002. 
We refer to this work as INN in this paper. 
In the INN approach, 
one of the conditions is given by the distance-redshift relation as in the MHE approach, 
and simultaneous big-bang or a uniform curvature function 
is chosen as the other condition. 
However, they could not go beyond $z\sim1.6$ owing to a technical problem. 
In 2008, Yoo, Kai and Nakao 
succeeded in 
constructing the LTB model whose distance-redshift relation
agrees with that of the concordance $\Lambda$CDM model 
in the whole redshift domain and which has uniform 
big-bang time~\cite{Yoo:2008su}. 
Independently from MHE and INN, 
C\'{e}l\'{e}rier solved the inverse problem 
analytically at small redshifts $z\ll1$
in the form of the Maclaurin series
in 1999~\cite{Celerier:1999hp}. 
There are also several works on 
the inverse problem~\cite{Chung:2006xh,Vanderveld:2006rb,Romano:2009mr,Romano:2009ej}.

In the inverse problem, one of the difficulties 
happens at the point with the maximum angular diameter distance. 
At this point, differential equations 
for the inverse problem become apparently singular. 
Related discussions can be found in 
Refs. \citen{Mustapha:1998jb,McClure:2007hy,
Hellaby:2006cj,Lu:2007gr,Celerier:2009sv}. 
The point with the maximum angular diameter distance 
is a regular singular point 
of the differential equations in the INN approach. 
In Ref. \citen{Yoo:2008su}, 
this singularity has been resolved 
using a shooting method 
to solve the 
differential equations for the inverse problem. 
On the other hand, 
there is no conclusive illustration on 
how to resolve this apparent singularity in the MHE approach. 
The main purpose in this paper is to explicitly 
show how this singularity can or cannot be resolved 
in the MHE approach.

In this paper, we solve the inverse problem 
by the MHE approach using a 
different formulation from MHE. 
We find that, for general functional forms of $\da(z)$ and $\mu(z)$, 
the regular solution does not necessarily exist in the whole 
redshift domain as pointed out in Ref. \citen{Mustapha:1998jb}. 
Then, we represent the necessary and sufficient condition 
for the existence of the regular solution 
in terms of $\da(z)$ and $\mu(z)$\cite{Mustapha:1998jb}. 
We also show that this condition is satisfied 
by the distance-redshift relation and 
the redshift-space mass density in $\Lambda$CDM models. 
Deriving regular differential equations for the inverse problem 
with the distance-redshift relation and redshift-space mass density 
in $\Lambda$CDM models, we numerically solve them 
for the case $(\Omega_{\rm M0},\Omega_{\Lambda0})=(0.3,0.7)$ 
and compare our result with those in 
previous works\cite{Celerier:2009sv,Kolb:2009hn,Dunsby:2010ts}. 
We propose a set of analytic fitting functions for the
resultant LTB universe model. 
We also explain how to solve the inverse problem by the INN approach 
with simultaneous big-bang in Appendix \ref{sec:inn}. 
We use the unit system given by $c=G=H_0=1$ throughout this paper, 
where $c$, $G$ and $H_0$ are the speed of light, 
Newton's constant and Hubble constant, respectively. 

\section{Equations}
\subsection{LTB dust universe}
As mentioned in the introduction,
we consider a spherically symmetric 
inhomogeneous universe filled with dust. 
This universe is described by an exact solution of 
the Einstein equations, known as the LTB solution. 
The metric of the LTB solution is given by
\begin{equation}
ds^2=-dt^2+\frac{\left(\partial_r R(t,r)\right)^2}{1-k(r)r^2}dr^2
+R^2(t,r)d\Omega^2, \label{eq:metric}
\end{equation}
where $k(r)$ is an arbitrary function of the radial coordinate $r$. 
The matter is dust whose stress-energy tensor is given by
\begin{equation}
T^{\mu\nu}=\rho u^\mu u^\nu,
\end{equation}
where $\rho=\rho(t,r)$ is the 
mass density, 
and $u^a$ is the four-velocity of 
the fluid element. 
The coordinate system in Eq.~(\ref{eq:metric}) is chosen
in such a way that $u^\mu=(1,0,0,0)$.

The area radius $R(t,r)$ 
satisfies one of the Einstein equations, 
\begin{equation}
(\delt R)^2=-k(r)r^2+\frac{2M(r)}{R}=-k(r)r^2+\frac{m(r)r^3}{3R}, 
\label{eq:Eeq1}
\end{equation}
where $M(r)$ is an arbitrary function related to 
the 
mass 
density $\rho$ by
\begin{equation}
4\pi \rho=\frac{\delr M}{R^2\delr R}=\frac{\frac{1}{2}r^2m(r)
+\frac{r^3}{6}\delr m(r)}{R^2\delr R}
\label{eq:Eeq2}
\end{equation}
and we have defined 
\begin{equation}
M(r)=\frac{m(r)r^3}{6}. 
\end{equation}

Following Ref. \citen{Tanimoto:2007dq}, we write the solution 
of Eq.~(\ref{eq:Eeq1}) in the form,
\begin{eqnarray}
R(t,r)&=&(6M(r))^{1/3}(t-\tb(r))^{2/3}S(x)
=rm(r)^{1/3}(t-\tb(r))^{2/3}S(x), 
\label{eq:YS}\\
x&=&k(r)r^2\left(\frac{t-\tb(r)}{6M(r)}\right)^{2/3}
=k(r)m(r)^{-2/3}(t-\tb(r))^{2/3}, \label{eq:defx}
\end{eqnarray}
where $\tb(r)$ is an arbitrary function 
that determines the big-bang time, 
and $S(x)$ is a function defined implicitly as
\begin{equation}
S(x)=
\left\{\begin{array}{lll}
\displaystyle
\frac{\cosh\sqrt{-\eta}-1}{6^{1/3}(\sinh\sqrt{-\eta}
-\sqrt{-\eta})^{2/3}}
\,;\qquad
&\displaystyle
x=\frac{-(\sinh\sqrt{-\eta}-\sqrt{-\eta})^{2/3}}{6^{2/3}}
\quad&\mbox{for}~~x<0\,,
\\
\displaystyle
\frac{1-\cos\sqrt{\eta}}{6^{1/3}(\sqrt{\eta}
-\sin\sqrt{\eta})^{2/3}}
\,;&\displaystyle
x=\frac{(\sqrt{\eta}-\sin\sqrt{\eta})^{2/3}}{6^{2/3}}
\quad&\mbox{for}~~x>0\,,
\end{array}\right.
\label{eq:defS}
\end{equation}
and $S(0)=({3}/{4})^{1/3}$. 
The function $S(x)$ is analytic for $x<(\pi/3)^{2/3}$. 
Some characteristics of the function $S(x)$ 
are given in Refs. \citen{Yoo:2008su} and \citen{Tanimoto:2007dq}.

\subsection{Basic equations}
As shown in the preceding subsection, 
the LTB solution has three arbitrary functions: 
$k(r)$, $m(r)$ and $t_{\rm B}(r)$. 
One of them is a gauge degree of freedom 
for rescaling of the radial coordinate $r$. 
We fix the remaining two functional degrees of 
freedom 
imposing the following physical conditions. 
\begin{itemize}
\item{The angular diameter distance is 
given by $D_{\rm A}(z)$ as 
a function of the redshift. }
\item{The redshift-space mass density is 
given by $\mu(z)$ as 
a function of the redshift.}
\end{itemize}

To determine $t_{\rm B}(r)$, $k(r)$ and $m(r)$ 
from the above conditions, we consider a past directed outgoing 
radial null geodesic that emanates from the 
observer at the center. 
This null geodesic is expressed in the form
\begin{eqnarray}
t&=&t(z),\\
r&=&r(z).
\end{eqnarray}
We assume that the 
observer is located at the symmetry center 
$r=0$ and observes the light ray at 
$t=t_0$. 
To fix the gauge freedom to rescale the radial coordinate $r$, 
we adopt the light-cone gauge condition such 
that the relation 
\begin{equation}
t=t_0-r
\end{equation}
is satisfied along the observed light ray. 

Then, basic equations to determine $t_{\rm B}$, $k$ and $m$ 
are 
given as follows: 
\begin{enumerate}
\item{\bf Null geodesic equations}

The null geodesic equations in the LTB solution 
are given as 
\begin{eqnarray}
(1+z)\frac{dt}{dz}
&=&-\frac{\partial _r R}{\partial_t\partial_rR},
\label{eq:nullgeo1}\\
(1+z)\frac{dr}{dz}
&=&\frac{\sqrt{1-k(r)r^2}}{\partial_t\partial_rR}. 
\label{eq:nullgeo2}
\end{eqnarray}
One of these equations can be derived from 
the other one and the null condition. 
Therefore, one of them is sufficient under 
the null condition. 

\item{\bf Null condition}

By virtue of the light-cone gauge condition, 
the null condition on the observed 
light ray takes the very simple form of 
\begin{equation}
\delr R=\sqrt{1-kr^2}. 
\label{eq:nullcon}
\end{equation}

\item{\bf Distance-redshift relation}

As mentioned, we assume that the angular diameter distance 
is given by $D_{\rm A}(z)$. 
Then, we have 
\begin{equation}
R=\da(z). 
\label{eq:discon}
\end{equation}

\item{\bf Redshift-space mass density}

We define the total mass of a shell between $z$ and $z+dz$ 
as 
\begin{equation}
\mathcal M:=4\pi \da^2\rho\frac{\delr R}{\sqrt{1-kr^2}}\frac{dr}{dz}dz. 
\end{equation}
Then, we consider the redshift-space mass density $\mu$ 
defined by
\begin{equation}
\mu=\frac{\mathcal M}{4\pi \da^2}
\Leftrightarrow\mu=\rho\frac{dr}{dz}. 
\end{equation}
As mentioned, we assume that 
the redshift-space mass density is given as a function 
of the redshift. 
Thus, we have  
\begin{equation}
\rho\frac{dr}{dz}=\mu(z). 
\label{eq:dencon}
\end{equation}

\end{enumerate}

\subsection{Rewriting the basic equations as differential equations}

We can regard $r(z)$, $m(z)$, $k(z)$ and $\tb(z)$ as 
mutually independent functions of the redshift $z$. 
Basic equations to determine these functions are \eqref{eq:nullgeo2}, 
\eqref{eq:nullcon}, \eqref{eq:discon} and \eqref{eq:dencon}. 

Combining Eqs. \eqref{eq:Eeq2} and \eqref{eq:dencon}, 
we find 
\begin{equation}
3m\frac{dr}{dz}+r\frac{dm}{dz}
=24\pi\mu\sqrt{1-kr^2}\frac{\da^2}{r^2}\equiv A, 
\label{eq:1-1}
\end{equation}
where we have used the relation
\begin{equation}
\delr m=\frac{dm/dz}{dr/dz}. 
\end{equation}

Differentiating Eq.\eqref{eq:YS} with $z$, we have 
\begin{equation}
B\frac{dr}{dz}+C\frac{dm}{dz}+D\frac{dk}{dz}
+E\frac{d\tau}{dz}=\frac{d\da}{dz}, 
\label{eq:1-2}
\end{equation}
where 
\begin{eqnarray}
B&=&m^{1/3}(t-\tb)^{2/3}S
-\frac{2}{3}rm^{1/3}(t-\tb)^{-1/3}(S+xS'), 
\label{eq:B}\\
C&=&\frac{1}{3}rm^{-2/3}(t-\tb)^{2/3}(S-2xS'), 
\label{eq:C}\\
D&=&rm^{-1/3}(t-\tb)^{4/3}S', 
\label{eq:D}\\
E&=&-\frac{2}{3}rm^{1/3}(t-\tb)^{-1/3}(S+xS'). 
\end{eqnarray}

From Eqs. \eqref{eq:YS} and \eqref{eq:nullcon}, we have 
\begin{equation}
F\frac{dr}{dz}-C\frac{dm}{dz}-D\frac{dk}{dz}-E\frac{d\tb}{dz}=0, 
\label{eq:1-3}
\end{equation}
where 
\begin{equation}
F=\sqrt{1-kr^2}-m^{1/3}(t-\tb)^{2/3}S. 
\label{eq:F}
\end{equation}

For $\delt\delr R$, we find the expression 
\begin{equation}
\delt\delr R=G\delr m+H\delr k+I\delr \tb+J,
\end{equation}
where 
\begin{eqnarray}
G&=&\delt C=\frac{1}{6}rm^{-2/3}(t-\tb)^{-1/3}\frac{1}{S^2}, 
\label{eq:G}\\
H&=&\delt D=\frac{2}{3}rm^{-1/3}(t-\tb)^{1/3}(2S'+xS''), 
\label{eq:H}\\
I&=&\delt E=\frac{1}{6}rm^{1/3}(t-\tb)^{-4/3}\frac{1}{S^2}, 
\label{eq:I}\\
J&=&\frac{2}{3}m^{1/3}(t-\tb)^{-1/3}(S+xS'). 
\label{eq:J}
\end{eqnarray}
Then, Eq. \eqref{eq:nullgeo2} can be rewritten as 
\begin{equation}
J\frac{dr}{dz}+G\frac{dm}{dz}+H\frac{dk}{dz}+I\frac{d\tb}{dz}
=\frac{\sqrt{1-kr^2}}{1+z}. 
\label{eq:1-4}
\end{equation}

Solving Eqs. \eqref{eq:1-1}, \eqref{eq:1-2}, \eqref{eq:1-3} and \eqref{eq:1-4} 
for $dr/dz$, $dm/dz$, $dk/dz$ and $d\tau/dz$, we have 
\begin{eqnarray}
\frac{dr}{dz}&=&\frac{1}{B+F}\frac{d\da}{dz}, 
\label{eq:2-1}\\
\frac{dm}{dz}&=&\frac{1}{r}\left(A-3m\frac{dr}{dz}\right)
=\frac{1}{r}\left(A-\frac{3m}{B+F}\frac{d\da}{dz}\right), 
\label{eq:2-2}\\
\frac{dk}{dz}&=&
\frac{2}{r}
\left\{\frac{\sqrt{1-kr^2}}{1+z}\tilE
-(F\tilI+\tilE J)\frac{dr}{dz}+r(
\tilC\tilI-\tilE\tilG)\frac{dm}{dz}
\right\}, 
\label{eq:2-3}\\
\frac{d\tb}{dz}&=&-\frac{2}{r}
\left\{\frac{\sqrt{1-kr^2}}{1+z}\tilD-
(F\tilH +\tilD J)\frac{dr}{dz}+r(\tilC\tilH-\tilD\tilG)\frac{dm}{dz}\right\}, 
\label{eq:2-4}
\end{eqnarray}
where $\tilC=C/r$, $\tilD=D/r$, $\tilE=E/r$, $\tilG=G/r$, 
$\tilH=H/r$ and $\tilI=I/r$, and we have used the relation
\begin{equation}
\tilD\tilI-\tilE\tilH=\frac{1}{2}. 
\end{equation}

\section{Solution near the center}

Before numerically solving Eqs. \eqref{eq:2-1} - \eqref{eq:2-4}, 
we need to identify the boundary conditions for the differential 
equations at the center. 
Since the center is the regular singular point of the differential 
equations, the boundary conditions must be appropriately chosen 
for a regular solution. 
For this purpose, we consider the Maclaurin series expansion of the 
functions near the center as follows:
\begin{eqnarray}
r&=&r_1z+\frac{1}{2}r_2z^2+\mathcal O(z^3),
\label{eq:exr1}\\
m&=&m_0+m_1z+\mathcal O(z^2),
\label{eq:exm1}\\
k&=&k_0+k_1z+\mathcal O(z^2),
\label{eq:exk1}\\
\tb&=&t_{\rm B1}z+\mathcal O(z^2). 
\label{eq:extb1}
\end{eqnarray}
Hereafter, each value at $(t,r)=(t_0,0)$ is denoted by 
the subscript $0$. 

As shown in Ref. \citen{Yoo:2008su}, 
imposing the regularity at the center, 
we find 
\begin{eqnarray}
R&=&r+\mathcal O(r^2), \\
m&=&m_0+\mathcal O(r), \\
m_0&=&8\pi\rho_0:=8\pi\rho(t_0,0), \\
k_0&=&x_0m_0^{2/3}t_0^{-2/3}=-1+\frac{m_0}{3}, \\
S_0&=&m_0^{-1/3}t_0^{-2/3}. 
\label{eq:fort0}
\end{eqnarray}
Since $S$ satisfies
\begin{eqnarray}
S+xS'&=&\frac{\sqrt{3}}{2}\sqrt{\frac{1}{S}-3x}
~~{\rm for}~~x\leq \left(\frac{\pi}{6}\right)^{2/3}, \\
S+xS'&=&\frac{\sqrt{3}}{2}\sqrt{\frac{1}{S}-3x}
~~{\rm for}~~x> \left(\frac{\pi}{6}\right)^{2/3}, 
\end{eqnarray}
we find 
\begin{equation}
S_0+x_0S_0'=\frac{3}{2}m_0^{-1/3}t_0^{1/3}, 
\end{equation}
where we have assumed
\begin{equation}
\Omega_{\rm m0}=\frac{m_0}{3}\leq1
\Leftrightarrow k_0\leq0
\Rightarrow x_0\leq0<\left(\frac{\pi}{6}\right)^{2/3}. 
\end{equation}

We assume the following expansion form of the angular diameter distance 
and the redshift-space mass density: 
\begin{eqnarray}
\da&=&z+\frac{1}{2}D_{\rm A2}z^2+\mathcal O(z^3),
\label{eq:daexpand}\\
8\pi\mu&=&\delta_0+\delta_1z+\mathcal O(z^2).
\end{eqnarray}
Then, from zero-th order of Eq. \eqref{eq:2-1}, we can find 
\begin{equation}
r_1=1. 
\end{equation}
Equation \eqref{eq:2-2} can be expanded as 
\begin{equation}
\frac{dm}{dz}=-3\frac{m_0-\delta_0}{z}+\mathcal O(1). 
\end{equation}
Therefore, we have to impose the condition
\begin{equation}
m_0=\delta_0
\end{equation}
so that $dm/dz$ is finite at the center. 
Assuming the above relation, we can find 
\begin{equation}
r_2=1+D_{\rm A2}
\end{equation}
from the second order in Eq. \eqref{eq:2-1}. 
In the same manner as in the above, we obtain 
\begin{eqnarray}
m_1
&=&-\frac{3}{4}\left(\left(2+D_{\rm A2}\right)\delta_0-\delta_1\right),\\
k_1
&=&2+D_{\rm A2}-\frac{1}{4}(2+D_{\rm A2})\delta_0+\frac{\delta_1}{4},\\ 
t_{\rm B1}
&=&\frac{-2 (3 t_0+2) 
\delta_0^2+3 (\delta_1+12) t_0 \delta_0-3 D_{\rm A2}
((\delta_0-6) t_0+2) \delta_0+18 \delta_1 
(t_0-1)}{8 (\delta_0-3)\delta_0} 
\end{eqnarray}
from the 1st orders of Eqs. \eqref{eq:2-2}--\eqref{eq:2-4}, 
where we have used the equation
\begin{equation}
x(2SS''+S'^2)+5SS'+\frac{9}{4}=0 
\end{equation}
to reduce the order of differentiation of $S$. 
Then, we can use these expressions near the center 
instead of solving Eqs. \eqref{eq:2-1}--\eqref{eq:2-4}. 
If $D_{\rm A}(z)$ and $\mu(z)$ are analytic 
at the center, we can derive 
higher order expressions for Eqs. \eqref{eq:exr1}--\eqref{eq:extb1} 
from Eqs. \eqref{eq:2-1}--\eqref{eq:2-4}.

It should be noted that 
all the initial valuables at the center 
are determined by the 
input valuables $\delta_0$ and $H_0=1$ 
from the regularity. 
We do not have any additional degree of freedom 
to put a boundary condition for the differential equations 
differently from the situation 
in Ref. \citen{Yoo:2008su}. 
We also note that the normalization $H_0=1$ implicitly determines the 
value of $t_0$ through Eq. \eqref{eq:fort0}.

\section{Regularity at the point with the maximum distance}

To obtain a physically reasonable solution, 
the right-hand side of Eq. \eqref{eq:2-1} must be 
positive definite. 
Since the sign of $d\da/dz$ changes at the 
point with the maximum angular diameter distance, 
at which $d\da/dz=0$, 
$B+F$ must be 0 at this point and the sign of $B+F$ must change. 
As shown in Appendix \ref{sec:derive}, we can derive the equation
\begin{equation}
B+F=\frac{1}{1+z}\left(1-Q(z)\right), 
\label{eq:bf}
\end{equation}
where 
\begin{equation}
Q(z)=4\pi\int^z_0(1+z)
\mu(z)\da(z) dz. 
\end{equation}
Therefore, the input functions $\da(z)$ and $\mu(z)$ 
must satisfy 
\begin{equation}
Q(z_{\rm m})=4\pi\int^{z_{\rm m}}_0(1+z)
\mu(z)\da(z) dz=1
\label{eq:regcond}
\end{equation}
for the regularity at $z=z_{\rm m}$, 
where $d\da/dz|_{z=z_{\rm m}}=0$. 
This condition has been pointed out in Ref.\citen{Mustapha:1998jb}. 


Let us consider the angular diameter distance and 
the redshift-space mass density in a homogeneous and isotropic universe 
model with $\Omega_{\rm M0}$, $\Omega_{\rm K0}$ and $\Omega_{\rm \Lambda 0}$. 
In this case, the angular diameter distance is given by 
\begin{equation}
\da(z)=D_{\Lambda {\rm CDM}}(z):=\left\{
\begin{array}{l}
\frac{1}{1+z}\int^z_0\frac{1}{H_{\Lambda {\rm CDM}}}dz~~{\rm for}~~\Omega_{K0}=0, \\
\frac{1}{(1+z)\sqrt{-\Omega_{\rm K0}}}
\sin\left[\sqrt{-\Omega_{\rm K0}}\int^z_0\frac{1}{H_{\Lambda {\rm CDM}}}dz\right]
~~{\rm for}~~\Omega_{K0}\neq0, 
\end{array}
\right. 
\end{equation}
where 
\begin{equation}
H_{\Lambda {\rm CDM}}
=\sqrt{(1+z)^3\Omega_{\rm M0}+(1+z)^2\Omega_{\rm K0}+\Omega_{\Lambda 0}}. 
\end{equation}
The redshift-space mass density is given by
\begin{equation}
\mu(z)=\mu_{\Lambda {\rm CDM}}(z):=
\frac{3}{8\pi}\Omega_{\rm M0}
\frac{(1+z)^3}{\sqrt{1+\Omega_{\rm K0}(1+z)^2D_{\Lambda{\rm CDM}}^2}}
\left(\frac{	dD_{\Lambda{\rm CDM}}}{dz}+\frac{D_{\Lambda{\rm CDM}}}{1+z}\right). 
\label{eq:Deltam}
\end{equation}
Then, the following equations are satisfied: 
\begin{eqnarray}
&&\frac{d}{dz}\left[(1+z)^2H_{\Lambda {\rm CDM}}\frac{dD_{\Lambda{\rm CDM}}}{dz}\right]
=-\frac{3}{2}\frac{(1+z)^3}{H_{\Lambda {\rm CDM}}}\Omega_{\rm M0}D_{\Lambda{\rm CDM}}, \\
&&\frac{d}{dz}\sqrt{1+\Omega_{\rm K0}(1+z)^2D_{\Lambda{\rm CDM}}^2}
=\frac{(1+z)}{H_{\Lambda {\rm CDM}}}\Omega_{\rm K0}D_{\Lambda{\rm CDM}}~~{\rm for}~~\Omega_{\rm K0}\neq0, \\
&&\frac{d}{dz}\left[(1+z)^2D_{\Lambda{\rm CDM}}^2\right]=\frac{(1+z)}{H_{\Lambda {\rm CDM}}}D_{\Lambda{\rm CDM}}
~~{\rm for}~~\Omega_{\rm K0}=0. 
\end{eqnarray}
Using these equations, we can find
\begin{eqnarray}
Q(z)&=&4\pi\int^{z}_0(1+z)
\mu_{\Lambda {\rm CDM}}(z)D_{\Lambda{\rm CDM}}(z) dz
\nonumber\\
&=&\int^{z}_0
\frac{3}{2}\Omega_{\rm M0}
(1+z)^2\Omega_{\rm K0}^{-1}\frac{d}{dz}\sqrt{1+\Omega_{\rm K0}(1+z)^2D_{\Lambda{\rm CDM}}^2}
\nonumber\\
&=&\int^{z}_0-\frac{d}{dz}\left[(1+z)^2H_{\Lambda {\rm CDM}}\frac{dD_{\Lambda{\rm CDM}}}{dz}\right]dz
=1-(1+z)^2H_{\Lambda {\rm CDM}}\frac{dD_{\Lambda {\rm CDM}}}{dz}
\end{eqnarray}
for $\Omega_{\rm K0}\neq0$. 
We can also derive the same equation for the case $\Omega_{\rm K0}=0$, 
as pointed out in Ref. \citen{Kolb:2009hn}. 
Therefore, the condition \eqref{eq:regcond} is automatically satisfied 
for homogeneous and isotropic universes with 
$\Omega_{\rm M0}$, $\Omega_{\rm K0}$ and $\Omega_{\rm \Lambda 0}$. 
In this case, we have 
\begin{equation}
B+F=(1+z)H_{\Lambda {\rm CDM}}\frac{dD_{\Lambda {\rm CDM}}}{dz}. 
\end{equation}
Then, Eq. \eqref{eq:2-1} can be replaced by 
\begin{equation}
\frac{dr}{dz}=\frac{1}{(1+z)H_{\Lambda {\rm CDM}}}. 
\label{eq:2-1i}
\end{equation}
The set of differential equations 
\eqref{eq:2-1i}, \eqref{eq:2-2}--\eqref{eq:2-4} does not have any 
singular point other than the center. 

\begin{figure}[htbp]
\begin{center}
\includegraphics[scale=]{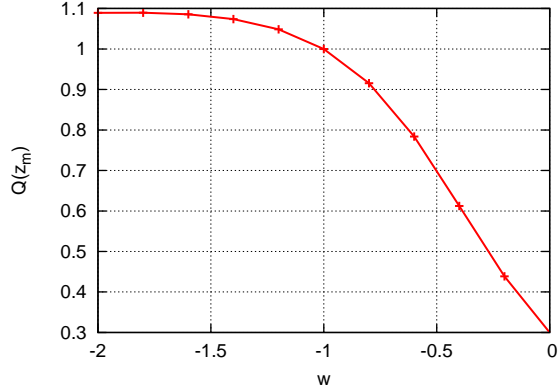}
\caption{$Q(z_{\rm m})$ as a function of $w$. 
$\da$ and $\mu$ are given by 
those in homogeneous and isotropic universe models with 
$\Omega_{\rm M0}=0.3$ and $\Omega_{\rm X0}=0.7$. 
}
\label{fig:qm}
\end{center}
\end{figure}

On the other hand, if we consider a dark energy component with 
the equation of state $p=w\rho~(w\neq-1)$ instead of 
the cosmological constant $\Lambda$, 
the condition \eqref{eq:regcond} cannot be satisfied. 
We numerically calculated the value of $Q(z_{\rm m})$, 
with $\Omega_{\rm M0}=0.3$ and 
$\Omega_{X0}=0.7$, where $\Omega_{\rm X0}$ is the 
density parameter of the dark energy component. 
The result is shown in Fig. \ref{fig:qm}. 
This result shows that 
there is no regular solution for 
the inverse problem with $\da$ and $\mu$ for 
a homogeneous universe model with a dark energy component with 
$w\neq1$ in the whole redshift domain. 
In other words, the existence of the regular solution for the 
inverse problem cannot be guaranteed for general input functions
$\da(z)$ and $\mu(z)$ without 
the condition \eqref{eq:regcond} 
as pointed out in Ref. \citen{Mustapha:1998jb}.

\section{Numerical results}
\label{sec:numeres}

\subsection{Numerical integration}
In this subsection, we consider 
the distance-redshift relation $\da$ and 
the redshift-space mass density $\mu$ for a 
homogeneous and isotropic universe with 
$(\Omega_{\rm M0},\Omega_{\rm X0})=(0.3,0.7)$, 
where $\Omega_{\rm X0}$ is the density parameter for 
the dark energy component whose equation of state 
is given by $p=w\rho$. 
The distance-redshift relation satisfies the following 
second-order differential equation and 
boundary conditions\cite{1988A&A...206..190L,Sereno:2001cc}:
\begin{equation}
\frac{d^2\da}{dz^2}=-\left(\frac{2}{1+z}+\frac{dH/dz}{H}\right)
\frac{d\da}{dz}-\frac{(1+z)\da}{2H^2}\left(
3\Omega_{\rm M0}+(1+w)\Omega_{\rm X0}(1+z)^{w-2}\right),
\end{equation}
\begin{equation}
\da(0)=0~~~{\rm ,}~~~\left.\frac{d\da}{dz}\right|_{z=0}=1, 
\end{equation}
where 
\begin{equation}
H=\sqrt{(1+z)^3\Omega_{\rm M0}+(1+z)^{w+1}\Omega_{\rm X0}}. 
\end{equation}
The redshift-space mass density is given by the same 
expression as Eq. \eqref{eq:Deltam}.

\begin{figure}[htbp]
\begin{center}
\includegraphics[scale=1.1]{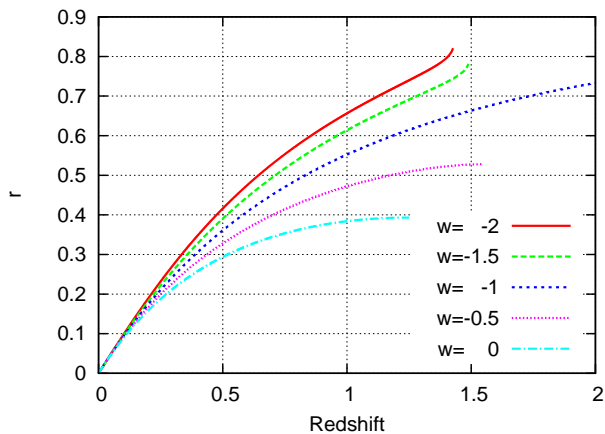}
\caption{Radial coordinate along the past directed null geodesic 
as a function of redshift for each value of $w$. 
}
\label{fig:rx-z}
\end{center}
\end{figure}

\begin{figure}[b]
\begin{center}
\includegraphics[scale=0.78]{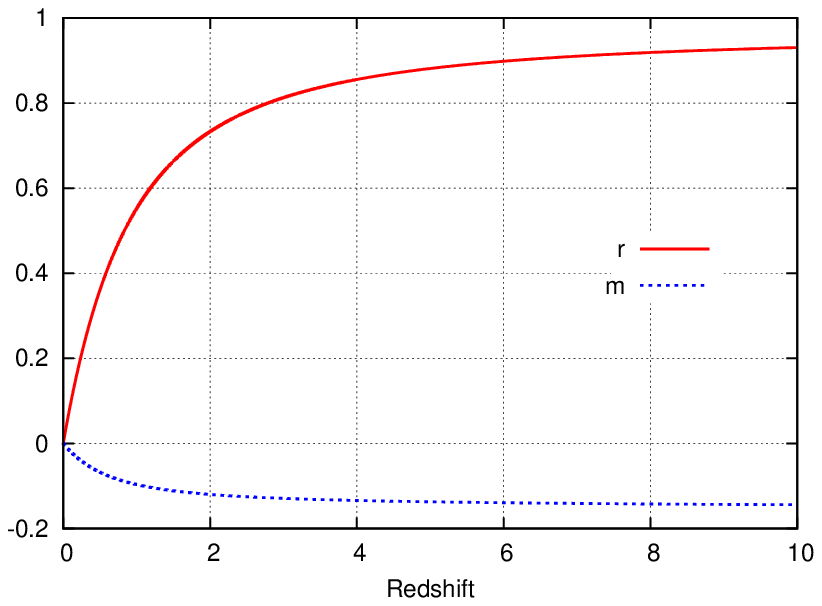}
\includegraphics[scale=0.78]{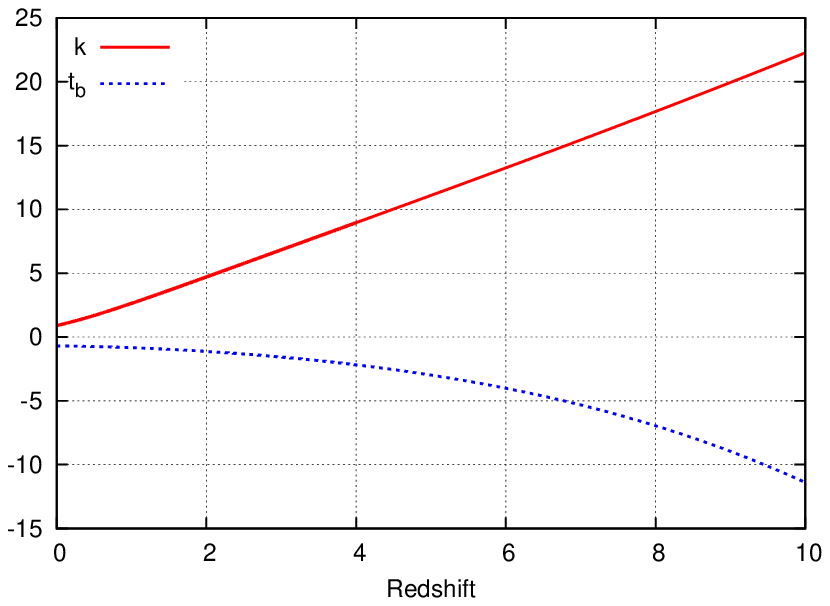}
\caption{$r$, $k$, $m$ and $\tb$ as functions of $z$. 
}
\label{fig:solsz}
\end{center}
\end{figure}

\begin{figure}[htbp]
\begin{center}
\includegraphics[scale=0.78]{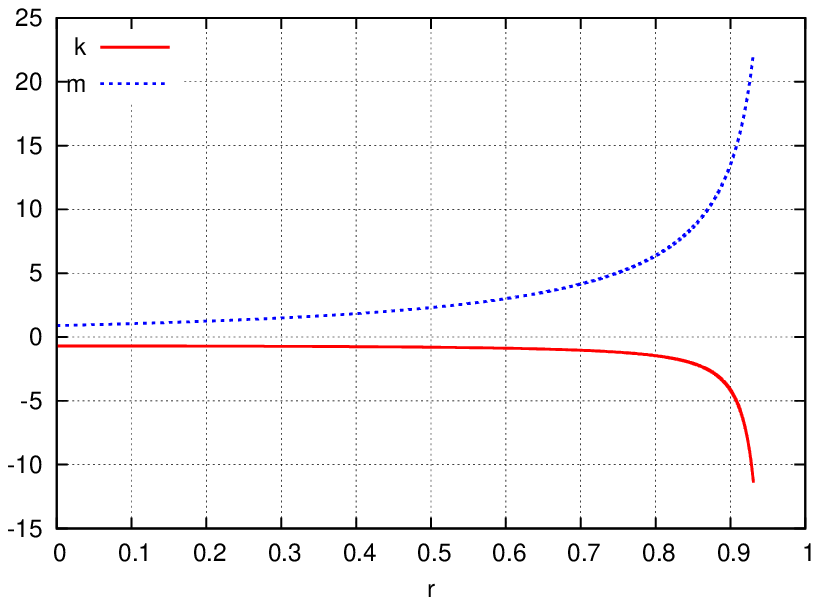}
\includegraphics[scale=0.78]{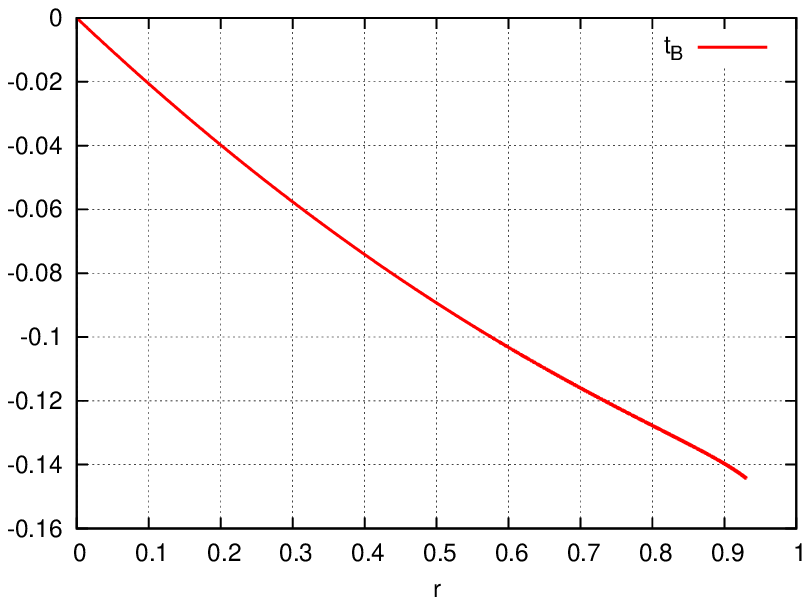}
\caption{$k$, $m$ and $\tb$ as functions of $r$. 
}
\label{fig:solsr}
\end{center}
\end{figure}

\begin{figure}[htbp]
\begin{center}
\includegraphics[scale=1]{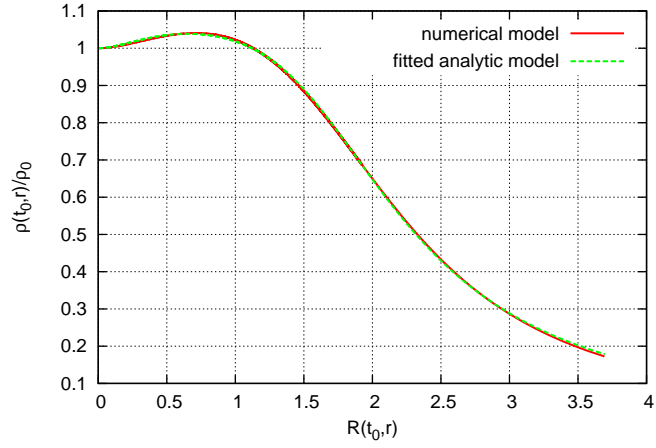}
\caption{$\rho(t_0,r)$ as a function of $R(t_0,r)$ for 
the numerically obtained model and the model with 
analytic fitting functions. 
}
\label{fig:rho0}
\end{center}
\end{figure}

We numerically integrated Eqs. \eqref{eq:2-1}--\eqref{eq:2-4}. 
Equation \eqref{eq:2-1i} is used 
instead of Eq. \eqref{eq:2-1} for the case $w=-1$. 
Near the center, 
solutions can be given in forms of the Maclaurin series, as 
shown in Appendix \ref{sec:expand}. 
We use these expressions near the center 
instead of solving the differential equations. 
As shown in Fig. \ref{fig:rx-z}, 
we can find a regular solution in the whole redshift domain 
only for $w=-1$. 
For $w=-1$, 
$r(z)$, $m(z)$, $k(z)$ and $t_{\rm B}(z)$ are 
depicted in Fig. \ref{fig:solsz}. 
$m$, $k$ and $t_{\rm B}$ are depicted as functions of 
$r$ in Fig. \ref{fig:solsr}. 
The energy density $\rho$ on the $t=t_0$ time slice 
is depicted as a function of circumferential radius $R(t_0,r)$ in 
Fig. \ref{fig:rho0}. 
This hump type profile has already been found in Ref. \citen{Celerier:2009sv} and 
consistent results can be seen in Refs. \citen{Kolb:2009hn} and \citen{Dunsby:2010ts}. 
Our result for $w=-1$ is also consistent with the result 
in these previous papers. 

\subsection{Gauge transformation and fitting functions}

We propose fitting functions for the given solution. 
Before that, for convenience, we perform a gauge transformation 
$\widetilde r=\widetilde r(r)$. 
Hereafter, the ``$\widetilde ~$" represents the quantities in the new gauge. 
We consider the new gauge defined by 
\begin{equation}
M(r)=\widetilde M(\widetilde r)=\frac{4\pi}{3}\widetilde r^3\rho_0. 
\end{equation}
Then, the relation between $r$ and $\widetilde r$ is given by 
\begin{equation}
\widetilde r^3=\frac{m(r)}{m_0}r^3. 
\end{equation}
The curvature function $\widetilde k(\widetilde r)$ and 
the big-bang time $\widetilde\tb(\widetilde r)$ are given by 
\begin{eqnarray}
&&\widetilde k(\widetilde r)\widetilde r^2=k(r)r^2,\\
&&\widetilde \tb(\widetilde r)=\tb(r). 
\end{eqnarray}

\begin{figure}[htbp]
\begin{center}
\includegraphics[scale=0.78]{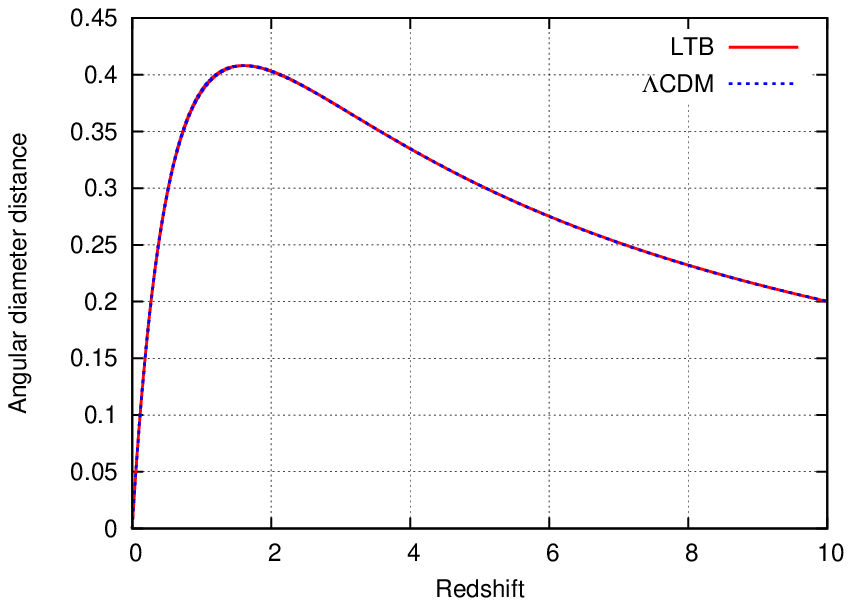}
\includegraphics[scale=0.78]{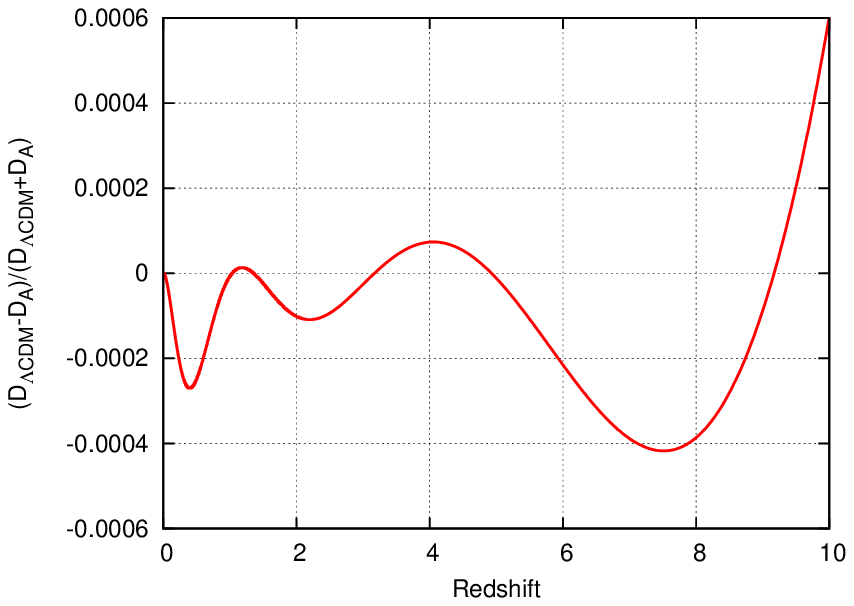}

\includegraphics[scale=0.78]{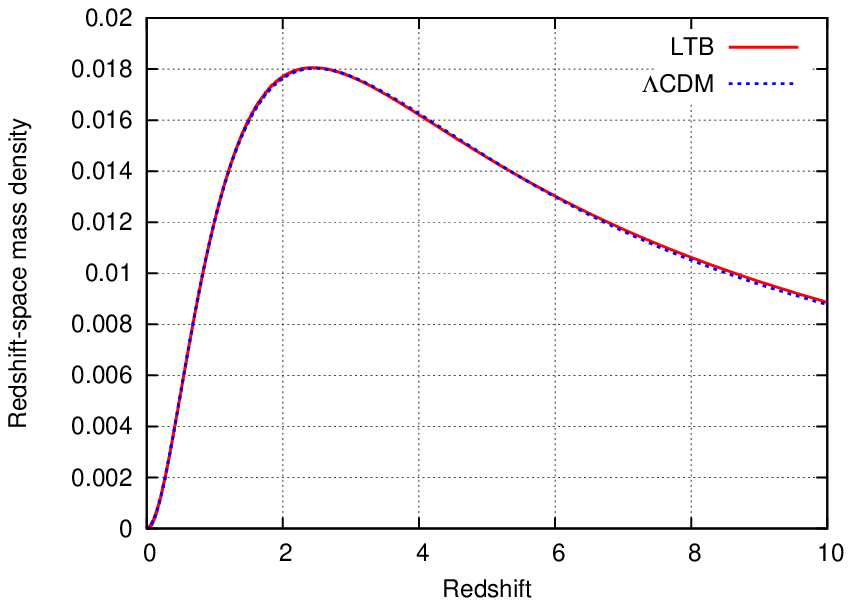}
\includegraphics[scale=0.78]{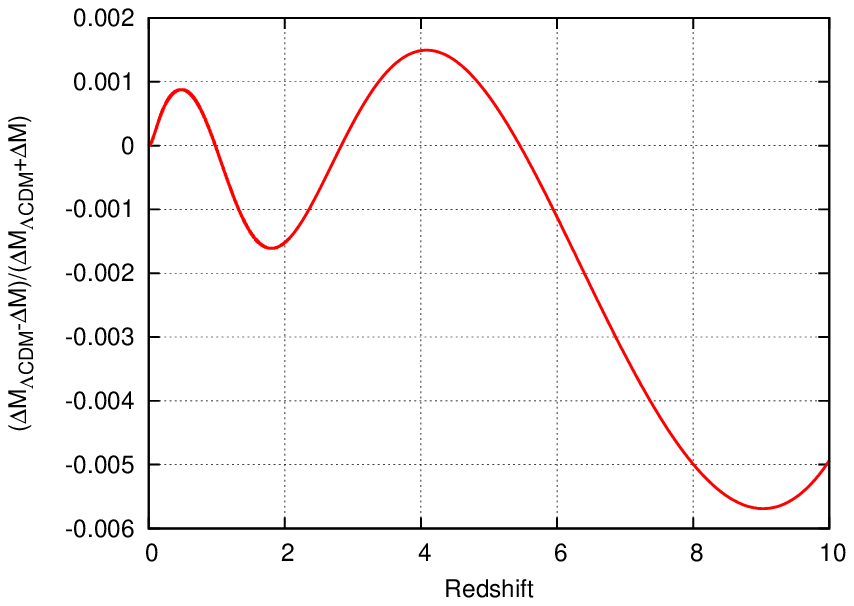}
\caption{Differences in the angular diameter distance and 
redshift-space mass density between the 
$\Lambda $CDM model and LTB model 
with $\widetilde k(\widetilde r)=\widetilde k_{\rm fit}(\widetilde r)$ and 
$\widetilde t_{\rm B}(\widetilde r)=\widetilde t_{\rm Bfit}(\widetilde r)$. 
}
\label{fig:difs}
\end{center}
\end{figure}

We propose analytic fitting functions for 
$\widetilde k$ and $\widetilde \tb$ 
as 
\begin{eqnarray}
\widetilde k_{\rm fit}(\widetilde r)&=&
-\frac{7}{10}
+0.728893 \widetilde r
-0.634917 \widetilde r^2
+0.303959 \widetilde r^3
-0.073768 \widetilde r^4
-\frac{0.00878216\widetilde r}{\widetilde r+0.303959}, ~~\\
\widetilde t_{\rm Bfit}(\widetilde r)&=&
-0.21319 \widetilde r
-0.013605 \widetilde r^2
+0.000925931 \widetilde r^3
+\frac{0.298252 \widetilde r^2}{\widetilde r+1.5439}. 
\end{eqnarray}
\begin{figure}[htbp]
\begin{center}
\includegraphics[scale=0.9]{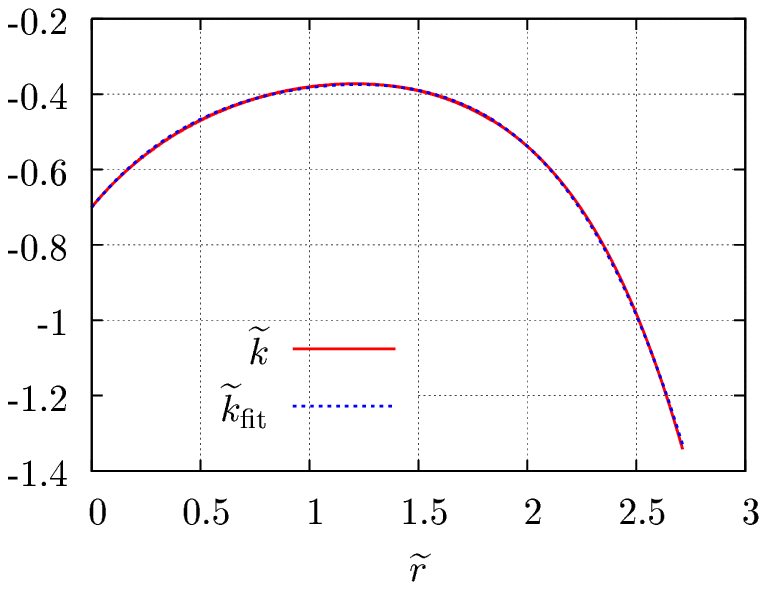}
\includegraphics[scale=0.9]{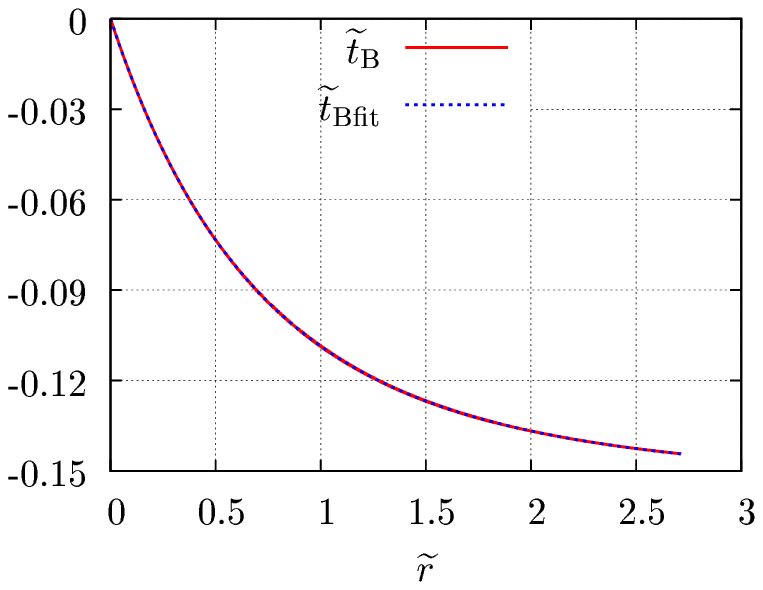}
\caption{$\widetilde k(\widetilde r)$, $\widetilde k_{\rm fit}(\widetilde r)$, 
$\widetilde t_{\rm B}(\widetilde r)$
 and $\widetilde t_{\rm Bfit}(\widetilde r)$. 
}
\label{fig:tildes}
\end{center}
\end{figure}
The differences in the angular diameter distance and 
redshift-space mass density between the 
$\Lambda $CDM model and LTB model 
with $\widetilde k(\widetilde r)=\widetilde k_{\rm fit}(\widetilde r)$ and 
$\widetilde t_{\rm B}(\widetilde r)=\widetilde t_{\rm Bfit}(\widetilde r)$ 
are less than 1\%, as shown in Fig. \ref{fig:difs}. 

2/15

\section{Summary}

As part of the summary, we 
describe a {\em theorem} and two {\em corollaries} based on 
the analysis in the previous sections after 
giving three {\em definitions} for convenience. 

\medskip
\underline{\em Definition 1}

We say that $\da(z)$ is {\em reasonable} as a function 
for the distance-redshift relation if it satisfies 
the following:
\begin{itemize}
\item{$\da(z)$ is a $C^1$ function 
on $z> 0$ and $\frac{d\da}{dz}$ is piecewise smooth on $z>0$. }
\item{$\da(z)>0$ for $z>0$ and $\da(0)=0$. }
\item{$\left.\frac{d\da}{dz}\right|_{z=0}=1/H_0=1$. }
\item{$\frac{d\da}{dz}=0$  at $z=z_{\rm m}$. }
\item{$\da(z)$ is $C^2$ and $\frac{d^2\da}{dz^2}<0$ at $z=z_{\rm m}$} 
\item{$\frac{d\da}{dz}\neq0$ for $z\neq z_{\rm m}$. }
\end{itemize}

\medskip
\underline{\em Definition 2}

We say that $\mu(z)$ is {\em reasonable} as 
a function for 
the redshift-space mass density if it satisfies 
the following:
\begin{itemize}
\item{$\mu(z)$ is finite and positive definite.}
\item{$\mu(z)$ is a piecewise smooth function 
on $z> 0$. }
\end{itemize}

\medskip
\underline{\em Definition 3}

We say that an LTB universe model is 
{\em observationally regular} 
if the LTB universe model is regular on the past light-cone of 
the observer at the center except for the big-bang initial singularity.

\medskip
\underline{\em Theorem}

For a set of a 
reasonable angular diameter distance $\da(z)$ and 
a reasonable redshift-space mass density $\mu(z)$, 
there exists the observationally regular LTB universe model 
whose distance-redshift relation and redshift-space mass density 
for the observer at the center agree with $\da(z)$ and 
$\mu(z)$, respectively, 
if and only if $\da(z)$ and $\mu(z)$ satisfy 
\begin{equation}
4\pi\int^{z_{\rm m}}_0(1+z)
\mu(z)\da(z) dz=1. 
\label{eq:regconda}
\end{equation}

\underline{\em Proof}

Since 
$$
B+F=\frac{1}{1+z}\left(1-4\pi\int^{z}_0(1+z)
\mu(z)\da(z) dz\right),
$$
Eq. \eqref{eq:regconda} is equivalent to 
$B+F=0$. 
We can find that $(1+z)(B+F)$ is a monotonically decreasing function of $z$, 
and $B+F$ can vanish only once in $z>0$. 
If $B+F\neq0$ at $z=z_{\rm m}$, 
since $\left.d\da/dz\right|_{z=z_{\rm m}}=0$, 
we have $\left.dr/dz\right|_{z=z_{\rm m}}=0$ 
from Eq. \eqref{eq:2-1}. 
Then, from Eq. \eqref{eq:dencon}, 
we find that $\rho$ must be infinite at $z=z_{\rm m}$ 
for a finite value of $\mu(z_{\rm m})$. 
Therefore, $B+F$ must vanish at $z=z_{\rm m}$ for 
the existence of the regular solution. 
That is, Eq. \eqref{eq:regconda} 
is a necessary condition for the existence of the 
regular solution. 
If the condition \eqref{eq:regconda} 
is satisfied, 
from L'H\^opital's rule, we have
\begin{equation}
\left.\frac{dr}{dz}\right|_{\rm z=z_{\rm m}}=
-\left.\frac{d^2\da/dz^2}{4\pi\mu(z) \da(z)}\right|_{\rm z=z_{\rm m}}>0.
\end{equation}
Then, we can find an observationally regular
LTB solution solving Eqs. \eqref{eq:2-1}--\eqref{eq:2-4}. \hfill Q.E.D.

\medskip

\medskip
\underline{\em Corollary-1}

For an observationally regular LTB universe model 
in which $\mu(z)$ and $\da(z)$ are reasonable,  
the distance-redshift relation $\da(z)$ and $\mu(z)$ 
satisfy $4\pi\int^{z_{\rm m}}_0(1+z)
\mu(z)\da(z) dz=1$. 

\underline{\em Proof}

Since Eqs. \eqref{eq:2-1}--\eqref{eq:2-4} are applicable for any 
LTB universe model, 
$4\pi\int^{z_{\rm m}}_0(1+z)
\mu(z)\da(z) dz=1$ 
must be satisfied so that an LTB universe model is 
observationally regular.  \hfill Q.E.D.

\medskip
\underline{\em Corollary-2}

There exist observationally regular LTB universe models 
whose distance-redshift relation 
and redshift-space mass density coincide with 
those in $\Lambda$CDM models.

\underline{\em Proof}

As shown in the text, $D_{\Lambda{\rm CDM}}(z)$ and $\mu_{\Lambda{\rm CDM}}(z)$ satisfy 
the condition \eqref{eq:regconda}. 
Then, this corollary is an immediate consequence from the above theorem.  \hfill Q.E.D.

\medskip

The condition \eqref{eq:regconda} has been derived 
in Ref. \citen{Mustapha:1998jb} at the first time. 
Without this condition, we cannot obtain 
the LTB universe model beyond $z=z_{\rm m}$. 
We demonstrated this using 
$\da(z)$ and $\mu(z)$ for flat 
homogeneous and isotropic universe models 
with the dark energy component whose equation of state 
is given by $p=w\rho$. 
If $w\neq-1$, we cannot obtain the regular LTB solution  
beyond $z=z_{\rm m}$ by solving the differential equations 
\eqref{eq:2-1}--\eqref{eq:2-4}.

In \S\ref{sec:numeres}, 
we have obtained the LTB universe model 
that realizes the distance-redshift relation and the redshift-space mass density 
in the $\Lambda$CDM model with 
$(\Omega_{\rm M0},\Omega_{\rm \Lambda0})=(0.3,0.7)$. 
Then, we introduced analytic fitting functions for this numerically obtained LTB model. 
The LTB model with these analytic fitting functions 
realize the distance-redshift relation and the redshift-space mass density 
in the $\Lambda$CDM model within 1\% accuracy. 

\section*{Acknowledgements}
We thank M. Sasaki, K. Nakao, T. Tanaka and T. Harada 
for valuable comments and useful suggestions.
C. Y. is supported by JSPS Grant-in-Aid for Creative Scientific Research No.~19GS0219.

\appendix
   
\section{Derivation of Eq. (\ref{eq:bf})}
\label{sec:derive}

Before the derivation, we 
provide some useful equations. 
\begin{itemize}
\item{
From the expression for $B+F$, we can find
\begin{equation}
B+F=-\delt R+\delr R. 
\label{eq:ap1}
\end{equation}
}
\item{
Differentiating Eq. \eqref{eq:Eeq1} with $t$, we have 
\begin{equation}
\delt^2R=-\frac{mr^3}{6R^2}. 
\label{eq:ap2}
\end{equation}
}
\item{Differentiating Eq. \eqref{eq:nullcon} with $z$ 
along the light-cone, we have 
\begin{equation}
\frac{d}{dz}\delr R=\frac{d}{dz}\sqrt{1-kr^2}
~~\Leftrightarrow~~\delr^2R=\delt\delr R+\delr \sqrt{1-kr^2}, 
\label{eq:ap3}
\end{equation}
where we have used the relation
\begin{equation}
\frac{d}{dz}=\frac{dr}{dz}\delr+\frac{dt}{dz}\delt
=\frac{dr}{dz}\left(\delr-\delt\right). 
\label{eq:ap4}
\end{equation}
}
\item{Differentiating Eq. \eqref{eq:Eeq1} with $r$ and 
dividing both sides by $\sqrt{1-kr^2}$, we have 
\begin{equation}
\delr \sqrt{1-kr^2}-\frac{mr^3}{6R^2}=\frac{\delt\delr R\delt R}{\sqrt{1-kr^2}}
-\frac{\delr\left(mr^3\right)}{6R\sqrt{1-kr^2}}. 
\label{eq:ap5}
\end{equation}
}
\end{itemize}
\medskip

First, 
differentiating Eq. \eqref{eq:ap1} with the redshift $z$, 
we have
\begin{equation}
\frac{d}{dz}(B+F)=\frac{dr}{dz}
\left(\delr^2R+\delt^2 R-2\delt \delr R\right). 
\end{equation}
Using Eqs. \eqref{eq:ap2} and \eqref{eq:ap3}, 
we can rewrite the above equation as 
\begin{equation}
\frac{d}{dz}(B+F)=\frac{dr}{dz}
\left(\delr\sqrt{1-kr^2}-\frac{mr^3}{6R^2}-\delt\delr R\right). 
\end{equation}
Then, using Eqs. \eqref{eq:ap5} and \eqref{eq:nullcon}, 
we have
\begin{eqnarray}
\frac{d}{dz}(B+F)&=&\frac{dr}{dz}\left(\frac{\delt\delr R}{\sqrt{1-kr^2}}
\left(\delt R-\delr R\right)
-\frac{\delr(mr^3)}{6R\sqrt{1-kr^2}}\right)\nonumber\\
&=&
\frac{dr}{dz}\left(-\frac{\delt\delr R}{\sqrt{1-kr^2}}
\left(B+F\right)
-\frac{\delr(mr^3)}{6R\sqrt{1-kr^2}}\right). 
\end{eqnarray}
Using Eqs. \eqref{eq:nullgeo2},
\eqref{eq:dencon} and \eqref{eq:2-1}, we can find
\begin{eqnarray}
&&\frac{d}{dz}(B+F)=-\frac{B+F}{1+z}-4\pi\mu \da \nonumber\\
&\Leftrightarrow&
\frac{d}{dz}\left((1+z)(B+F)\right)=-4\pi(1+z)\mu \da. 
\end{eqnarray}
Since $\left.(B+F)\right|_{z=0}=1$, we have 
\begin{equation}
B+F=\frac{1}{1+z}\left(1-4\pi\int^z_0(1+z)\mu\da dz\right). 
\end{equation}

\section{Series Expansion of the Solution}
\label{sec:expand}

We consider the distance-redshift relation $\da$ and the 
redshift-space mass density $\mu$ 
for homogeneous and isotropic universes with 
$(\Omega_{\rm M0},\Omega_{\rm X0},\Omega_{\rm K0}
=1-\Omega_{\rm M0}-\Omega_{\rm X0})$ 
as input functions 
for the inverse problem given by 
a set of equations, Eqs. \eqref{eq:2-1} - \eqref{eq:2-4}. 
For the solution near the center, 
we write functions $r(z)$, $m(z)$, $k(z)$ and 
$\tb(z)$ in the Maclaurin series as follows:
\begin{eqnarray}
r&=&r_1z+\frac{1}{2}r_2z^2+\frac{1}{6}r_3z^3+\frac{1}{24}r_4z^4
+\mathcal O(z^5),
\label{eq:exr2}\\
m&=&m_0+m_1z+\frac{1}{2}m_2z^2+\frac{1}{6}m_3z^3+\mathcal O(z^4),
\label{eq:exm2}\\
k&=&k_0+k_1z+\frac{1}{2}k_2z^2+\frac{1}{6}k_3z^3+\mathcal O(z^4),
\label{eq:exk2}\\
\tb&=&t_{\rm B1}z+\frac{1}{2}t_{\rm B2}z^2
+\frac{1}{6}t_{\rm B3}z^3+\mathcal O(z^4). 
\label{eq:extb2}
\end{eqnarray}
Substituting these expressions into 
Eqs. \eqref{eq:2-1}--\eqref{eq:2-4}, 
we can find the values of each coefficient as shown below.
For notational simplicity, we use $y:=3w+1$. 
\begin{eqnarray}
r(z)&=&z+\frac{1}{4} (-\Omega_{\rm M0}-\Omega_{\rm X0} y-4) z^2
   \nonumber\\
&+&\frac{1}{24}
   \left(3 \Omega_{\rm M0}^2+(6 \Omega_{\rm X0} y+8)
   \Omega_{\rm M0}+\Omega_{\rm X0} (y ((3 \Omega_{\rm X0}-2)
   y+8)-4)+24\right) z^3
      \nonumber\\
&+&\frac{1}{192} \bigl(-15 \Omega_{\rm M0}^3-9 (5
   \Omega_{\rm X0} y+4) \Omega_{\rm M0}^2-3 (\Omega_{\rm X0} (y (3 (5
   \Omega_{\rm X0}-2) y+26)-8)+24) \Omega_{\rm M0}
      \nonumber\\
&+&\Omega_{\rm X0} \left(y
   \left((3 (6-5 \Omega_{\rm X0}) \Omega_{\rm X0}-4) y^2-42 \Omega_{\rm X0}
   y+28 y+24 \Omega_{\rm X0}-92\right)+56\right)-192\bigr)
   z^4
      \nonumber\\
&+&O\left(z^5\right),
\end{eqnarray}
\begin{eqnarray}
m(z)&=&3 \Omega_{\rm M0}+\frac{9 \Omega_{\rm M0} z}{2}+\frac{3}{80}
   \Omega_{\rm M0} (-50 \Omega_{\rm M0}+24 \Omega_{\rm X0}+10
   \Omega_{\rm X0} y+60) z^2
         \nonumber\\
&+&\frac{3}{80} \Omega_{\rm M0} \bigl(25
   \Omega_{\rm M0}^2-12 \Omega_{\rm X0} \Omega_{\rm M0}+20 \Omega_{\rm X0} y
   \Omega_{\rm M0}-50 \Omega_{\rm M0}-5 \Omega_{\rm X0}^2 y^2+5
   \Omega_{\rm X0} y^2+22 \Omega_{\rm X0}
         \nonumber\\
&-&12 \Omega_{\rm X0}^2 y+20
   \Omega_{\rm X0} y+10\bigr) z^3+O\left(z^4\right),
   \end{eqnarray}
   \begin{eqnarray}
   k(z)&=&-1+\Omega_{\rm M0}+\left(\Omega_{\rm M0}-\frac{\Omega_{\rm X0}
   y}{2}-1\right) z
   \nonumber\\
&+&\frac{1}{240} \left(-100 \Omega_{\rm M0}^2+2
   (\Omega_{\rm X0} (25 y+36)+80) \Omega_{\rm M0}-5 \Omega_{\rm X0} (y
   ((8-3 \Omega_{\rm X0}) y+16)+16)-60\right) z^2
   \nonumber\\
&+&\frac{1}{480} \bigl(100
   \Omega_{\rm M0}^3+(\Omega_{\rm X0} (50 y-72)-200)
   \Omega_{\rm M0}^2
   \nonumber\\
&+&(\Omega_{\rm X0} (y (50 y-\Omega_{\rm X0} (65
   y+72)+190)+148)+100) \Omega_{\rm M0}
   \nonumber\\
&-&5 \Omega_{\rm X0} (y (y ((3
   (\Omega_{\rm X0}-2) \Omega_{\rm X0}+4) y+4)+24)+8)\bigr)
   z^3+O\left(z^4\right),
   \end{eqnarray}
   \begin{eqnarray}
   \tb(z)&=&-\frac{(\Omega_{\rm X0} (3 t_0-2) y) z}{4
   (\Omega_{\rm M0}-1)}
   \nonumber\\
&+&\frac{\Omega_{\rm X0} }{1920
   (\Omega_{\rm M0}-1)^3}\Bigl(-8 (-36 \Omega_{\rm M0}
   t_0+48 t_0-8) (\Omega_{\rm M0}-1)^2
   \nonumber\\
&-&10 (24 \Omega_{\rm X0}
   \Omega_{\rm M0}-18 \Omega_{\rm X0} t_0 \Omega_{\rm M0}+48
   t_0 \Omega_{\rm M0}-32 \Omega_{\rm M0}-84 \Omega_{\rm X0}
   \nonumber\\
&+&108
   \Omega_{\rm X0} t_0-48 t_0+32) y^2 (\Omega_{\rm M0}-1)
   \nonumber\\
&+&4
   \left(90 t_0 \Omega_{\rm M0}^2-60 \Omega_{\rm M0}^2+90 t_0
   \Omega_{\rm M0}-60 \Omega_{\rm M0}-180 t_0+120\right) y
   (\Omega_{\rm M0}-1)\Bigr)z^2
   \nonumber\\
&+&\frac{\Omega_{\rm X0}}{1920 (\Omega_{\rm M0}-1)^3} \Bigl(-5 \bigl(2 (12
   t_0+\Omega_{\rm X0} (-18 t_0+3 \Omega_{\rm X0} (3
   t_0-4)+20)-8) \Omega_{\rm M0}^2
   \nonumber\\
&+&\left((92-81 t_0)
   \Omega_{\rm X0}^2+32 (6 t_0-5) \Omega_{\rm X0}-48
   t_0+32\right) \Omega_{\rm M0}
   \nonumber\\
&+&2 (12 t_0+3 \Omega_{\rm X0}
   (-26 t_0+\Omega_{\rm X0} (28 t_0-23)+20)-8)\bigr) y^3
   \nonumber\\
&-&10
   (\Omega_{\rm M0}-1) (-24 \Omega_{\rm X0} \Omega_{\rm M0}^2+27
   \Omega_{\rm X0} t_0 \Omega_{\rm M0}^2-18 t_0
   \Omega_{\rm M0}^2+12 \Omega_{\rm M0}^2+30 \Omega_{\rm X0}
   \Omega_{\rm M0}
   \nonumber\\
&-&54 \Omega_{\rm X0} t_0 \Omega_{\rm M0}-30
   t_0 \Omega_{\rm M0}+20 \Omega_{\rm M0}+84 \Omega_{\rm X0}-108
   \Omega_{\rm X0} t_0+48 t_0-32) y^2
   \nonumber\\
&+&4
   (\Omega_{\rm M0}-1) (-45 t_0 \Omega_{\rm M0}^3+30
   \Omega_{\rm M0}^3-36 \Omega_{\rm X0} t_0 \Omega_{\rm M0}^2+120
   t_0 \Omega_{\rm M0}^2-40 \Omega_{\rm M0}^2-40 \Omega_{\rm X0}
   \Omega_{\rm M0}
   \nonumber\\
&+&126 \Omega_{\rm X0} t_0 \Omega_{\rm M0}-255
   t_0 \Omega_{\rm M0}+90 \Omega_{\rm M0}+60 \Omega_{\rm X0}-120
   \Omega_{\rm X0} t_0+180 t_0-80) y
   \nonumber\\
&-&8
   (\Omega_{\rm M0}-1)^2 \left(18 t_0 \Omega_{\rm M0}^2-9 t_0
   \Omega_{\rm M0}+18 \Omega_{\rm M0}-48 t_0+8\right)\Bigr)
   z^3+O\left(z^4\right). 
   \end{eqnarray}

\section{Inverse Problem with $t_{\rm B}(r)=0$}
\label{sec:inn}

Here, we show how to 
solve the inverse problem using the INN approach 
with $\tb(r)=0$ for a given 
angular diameter distance $\da(z)$. 
This has been carried out in Ref. \citen{Yoo:2008su} 
by solving a set of four differential equations 
parametrized by an affine parameter on the null geodesic. 
Here, we do not use the affine parameter but 
the redshift $z$ as the independent variable. 
In this procedure, the number of differential equations 
is reduced to three. 

\subsection{Basic equations}

Basic equations are derived by replacing 
the condition \eqref{eq:dencon} with $\tb(z)=0$. 
Then, we have the following three coupled first-order differential 
equations:
\begin{eqnarray}
\frac{dr}{dz}&=&\frac{1}{B+F}\frac{d\da}{dz}, 
\label{eq:tb0-1}\\
\frac{dm}{dz}&=&\frac{1}{DG - CH}\left(\frac{\sqrt{1 - kr^2}}{1 + z}D 
- (HF + DJ)\frac{dr}{dz}\right), 
\label{eq:tb0-2}\\
\frac{dk}{dz}&=&\frac{1}{D}\left(F\frac{dr}{dz} - C\frac{dm}{dz}\right). 
\label{eq:tb0-3}
\end{eqnarray}

\subsection{Solutions near the center}

We perform the Maclaurin series expansion as 
\eqref{eq:exr1}--\eqref{eq:daexpand}.  
Substituting these expressions into 
Eqs. \eqref{eq:tb0-1}--\eqref{eq:tb0-3}, 
we find  
\begin{eqnarray}
r(z)&=&z+\frac{1}{2} (D_{\rm A2}+1) z^2+\frac{1}{12} (4 D_{\rm A2}+2 D_{\rm A3}+m_0) z^3+O\left(z^4\right), 
   \label{eq:series-r}
\end{eqnarray}
\begin{eqnarray}
m(z)&=&m_0+\frac{m_0 (m_0+D_{\rm A2} (6-9 t_0)-18 t_0+9) z}{(m_0+6)
   t_0-6}+O\left(z^2\right), 
  \label{eq:series-m}
   \end{eqnarray}
\begin{eqnarray}
k(z)&=&\frac{m_0-3}{3}+\frac{(m_0-3) (m_0-6 D_{\rm A2} (t_0-1)-12 t_0+12) z}{3
   ((m_0+6) t_0-6)}+O\left(z^2\right). 
   \label{eq:series-k}
   \end{eqnarray}
Then, we find that the parameter $m_0$ can be freely chosen differently from 
the MHE approach.\footnote{We accept a nonvanishing first derivative of the 
density at the center\cite{Yoo:2008su,Vanderveld:2006rb}.}
We use this degree of freedom to guarantee 
$B+F=0$ at $z=z_{\rm m}$.

\subsection{Numerical procedure to solve the differential equations}

For illustrative purposes, we rewrite 
Eqs. \eqref{eq:tb0-1}--\eqref{eq:tb0-3} in abstract forms as 
\begin{eqnarray}
\frac{dr}{dz}&=&\frac{1}{X(z,r,m,k)}\frac{d\da}{dz}, 
\label{eq:tb0r-1}\\
\frac{dm}{dz}&=&Y\left(z,r,m,k,\frac{dr}{dz}\right), 
\label{eq:tb0r-2}\\
\frac{dk}{dz}&=&Z\left(z,r,m,k,\frac{dr}{dz},\frac{dm}{dz}\right). 
\label{eq:tb0r-3}
\end{eqnarray}
Then, the procedure to solve Eqs.\eqref{eq:tb0-1}-\eqref{eq:tb0-3}
is summarized in follows:
\begin{enumerate}
\item{We determine a trial value for $m_0$. }
\item{We solve the differential equations from the center, 
where, near the center, we use the expressions 
\eqref{eq:series-r}--\eqref{eq:series-k} 
instead of solving the differential equations. 
}
\item{We stop integrating the equations at $z=z_{\rm b}<z_{\rm m}$ and 
read off the values of $r$, $m$ and $k$ at this point.
We label these values as $r_{\rm b-}$, $m_{\rm b-}$ and $k_{\rm b-}$. }
\item{We determine trial values for $m_{\rm m}$ and $k_{\rm m}$, 
where the subscript ``m" denotes the value at $z=z_{\rm m}$.  }
\item{We numerically solve the equation 
$X(z_{\rm m},r_{\rm m}, m_{\rm m}, k_{\rm m})=0$
for $r_{\rm m}$. }
\item{At $z=z_{\rm m}$, Eq. \eqref{eq:tb0r-1} 
can be rewritten as 
\begin{equation}
\left.\frac{dr}{dz}\right|_{z=z_{\rm m}}
=\left.\frac{\frac{d^2\da}{dz^2}}{\frac{dX}{dz}(z,r,m,k,dr/dz,dm/dz,dk/dz)}
\right|_{z=z_{\rm m}}. 
\label{eq:drdzm}
\end{equation}
Then, Eqs. \eqref{eq:tb0r-2}--\eqref{eq:drdzm}  
can be regarded as simultaneous equations for 
$dr/dz|_{z=z_{\rm m}}$, $dm/dz|_{z=z_{\rm m}}$ and $dk/dz|_{z=z_{\rm m}}$. 
We numerically solve them. 
}
\item{We integrate the differential equations \eqref{eq:tb0-1}--\eqref{eq:tb0-3}
from $z=z_{\rm m}$ backward.} 
\item{We stop integrating the equations at $z=z_{\rm b}$ and read off 
the values of $r$, $m$ and $k$ at this point.
We label these values as $r_{\rm b+}$, $m_{\rm b+}$ and $k_{\rm b+}$.} 
\item{We define deviations $\delta r$, $\delta m$ and $\delta k$ as follows:
\begin{eqnarray}
\delta r&:=&r_{\rm b-}-r_{\rm b+}, \\
\delta m&:=&m_{\rm b-}-m_{\rm b+}, \\
\delta k&:=&k_{\rm b-}-k_{\rm b+}. 
\end{eqnarray}
 }
\item{
Operations 1--9 give the deviations $\delta r$, $\delta m$ and $\delta k$ 
for given values of $m_0$, $m_{\rm m}$ and $k_{\rm m}$. 
That is, we can define 
$\delta r=\delta r(m_0,m_{\rm m},k_{\rm m})$, 
$\delta m=\delta m(m_0,m_{\rm m},k_{\rm m})$ and 
$\delta k=\delta k(m_0,m_{\rm m},k_{\rm m})$ as functions of 
$m_0$, $m_{\rm m}$ and $k_{\rm m}$ using operations 1--9.
Then, repeating operations 1--9 and using the Newton-Raphson method, 
we numerically solve the equations 
\begin{eqnarray}
\delta r(m_0,m_{\rm m},k_{\rm m})&=&0, \\
\delta m(m_0,m_{\rm m},k_{\rm m})&=&0, \\
\delta k(m_0,m_{\rm m},k_{\rm m})&=&0
\end{eqnarray}
for $m_0$, $m_{\rm m}$ and $k_{\rm m}$. 
Eventually, we have functional forms 
of $r(z)$, $m(z)$ and $k(z)$ in $z<z_{\rm m}$ 
without discontinuity at $z=z_{\rm b}$. 
}
\item{Finally, we solve the differential equations 
\eqref{eq:tb0-1}--\eqref{eq:tb0-3}
from $z=z_{\rm m}$ forward with initial conditions 
$r(z_{\rm m})=r_{\rm m}$, $m(z_{\rm m})=m_{\rm m}$ and $k(z_{\rm m})=k_{\rm m}$ given above. }
\end{enumerate}

\subsection{Results}

\begin{figure}[htbp]
\begin{center}
\includegraphics[scale=0.78]{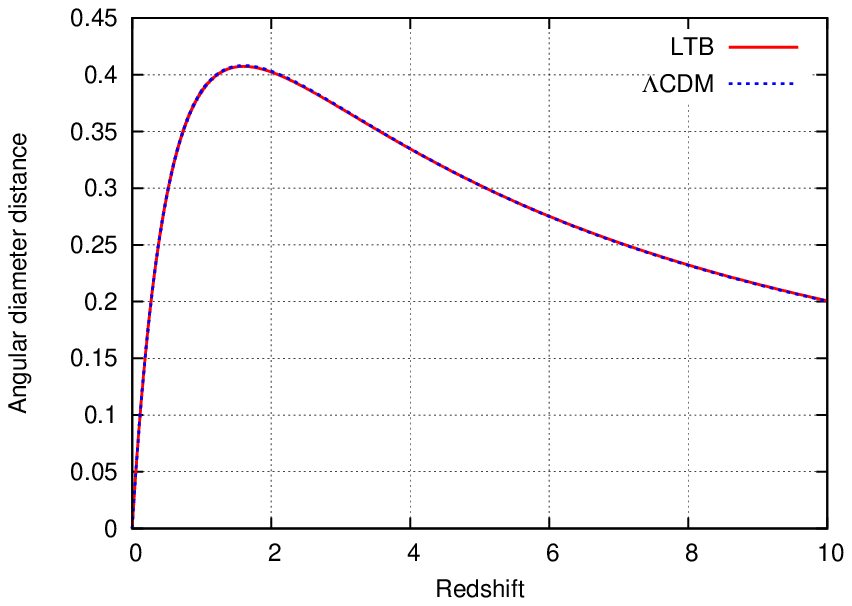}
\includegraphics[scale=0.78]{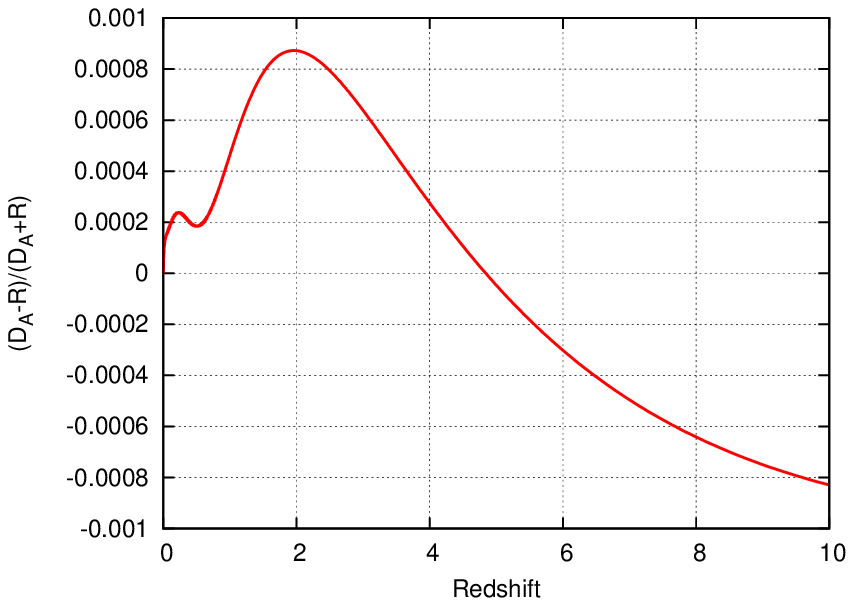}
\caption{Distance-redshift relations in the $\Lambda$CDM model 
with $(\Omega_{\rm M0},\Omega_{\Lambda 0})=(0.3,0.7)$ and 
the LTB model with 
$(\tilde t_{\rm B}(\tilde r),\tilde k(\tilde r))
=(0,\tilde k_{\rm fit}(\tilde r))$(left panel) 
and the deviation of the distance(right panel). 
}
\label{fig:difstb0}
\end{center}
\end{figure}

We performed operations 1--11 in the previous subsection using 
the distance-redshift relation in 
the $\Lambda$CDM model with $(\Omega_{\rm M0},\Omega_{\Lambda 0})=(0.3,0.7)$. 
The result is consistent with that in Ref. \citen{Yoo:2008su}. 
We performed the same gauge transformation as that 
given in \S\ref{sec:numeres}. 
Then, we find 
a fitting function for $\tilde k(\tilde r)$ 
as\cite{Yoo:2010qy}
\begin{equation}
\tilde k_{\rm fit}(\tilde r)
=\frac{0.545745}{0.211472+ \sqrt{0.026176+ \tilde r}} 
- \frac{2.22881}{\left(0.807782+ \sqrt{0.026176+ \tilde r}\right)^2}. 
\end{equation}
The angular diameter distance in 
the LTB universe model with $\tilde k(\tilde r)=\tilde k_{\rm fit}(\tilde r)$ 
is depicted in the Fig. \ref{fig:difstb0}. 
The deviation of the distance from that in the $\Lambda$CDM model is within 0.1\%. 


\end{document}